\documentstyle[prl,aps,preprint]{revtex}
\newcommand{\tr}{{\rm tr}}

\begin{document}

\draft
\tighten
\def\footnoterule{\kern-3pt \hrule width\hsize \kern3pt}

\title{Impediments to mixing classical and quantum dynamics}

\author{J. Caro and L.L. Salcedo}

\address{
{~} \\
Departamento de F\'{\i}sica Moderna \\
Universidad de Granada \\
E-18071 Granada, Spain
}

\date{\today}
\maketitle

\thispagestyle{empty}

\begin{abstract}
The dynamics of systems composed of a classical sector plus a quantum
sector is studied. We show that, even in the simplest cases, (i) the
existence of a consistent canonical description for such mixed systems
is incompatible with very basic requirements related to the time
evolution of the two sectors when they are decoupled. (ii) The
classical sector cannot inherit quantum fluctuations from the quantum
sector. And, (iii) a coupling among the two sectors is incompatible
with the requirement of physical positivity of the theory, i.e., there
would be positive observables with a non positive expectation value.
\end{abstract}



\section{Introduction}
\label{sec:1}

Ever since the beginnings of quantum mechanics, physical systems have
been considered which are composed of a quantum mechanical sector plus
another sector described in classical terms~\cite{Maddox:1995}. For
instance, this issue is central in the quantum theory of the
measurement when the apparatus is treated classically. The same
situation appears also at a less fundamental level. There are many
systems in the literature which are routinely treated using a mixed
quantum-classical description even if, as far as we know, they are
well accounted for by quantum mechanics. Molecular theory or quantum
optics are just two instances of this. The mixed description is used
as a convenient approximation which greatly simplifies the treatment
of such systems. In other cases some degrees of freedom are treated
classically because no complete quantum theory exist for them. A
typical example is the coupling of matter to gravity. In this case it
is standard to use a mean field treatment (called semiclassical
gravity) where a classical gravitational field obeys Einstein
equations using as source the expectation value of the energy-momentum
tensor of the quantum matter fields~\cite{Rosenfeld:1963}. When used
in early universe cosmology, this approach leads to universes which
are much too uniform as compared to present
observations~\cite{Brandenberger:1985cz}. This has been
attributed~\cite{Boucher:1988ua} to the fact that the mean field
approach misses the secondary quantum fluctuations induced on the
classical gravitational field by its coupling to the quantum matter
fields, the so-called quantum backreaction~\cite{Anderson:1995tn}. Of
course, in a full quantum treatment, the gravitational field would
present their own primary quantum fluctuations. Such a treatment has
not been pursued because of the lack of a renormalizable quantum
theory of gravitation. In principle one would expect that the
renormalizability problem would be less severe when the gravitational
field is classical although with quantum backreaction, however,
Ref.~\cite{Salcedo:1994sn} shows that very likely this will not be the
case. There, it is shown that removing the primary quantum
fluctuations of just one sector makes a formerly renormalizable theory
into a non renormalizable one.

The problem of mixing classical and quantum degrees of freedom has
been addressed by many authors from different points of view. A very
incomplete list is
~\cite{Boucher:1988ua,Anderson:1995tn,Salcedo:1994sn,DeWitt:1962,Aleksandrov:1981,Jones:1994,Salcedo:1996jr,Diosi:1996,Prezhdo:1997gs,Diosi:1997mt}.
There is no generally accepted definition of what is meant by a
classical-quantum system. This is natural since, as far as we know, no
such system exits in nature. In the present work a ``quantum-classical
mixing'' will mean some limit case of a quantum system. The issue that
we want to study is whether such a limit can actually be taken in a
way which is universal and internally consistent. Universal here
refers to the existence of a well-defined set of rules to be applied
to any quantum-quantum system to obtain its classical-quantum
version. Since our mixed systems are just degenerated cases of quantum
systems, no new universal parameters should be introduced and thus our
conclusions do not directly apply to approaches such as that in
\cite{Diosi:1997mt}.

As is well known, the Poisson bracket, which governs the classical
dynamics, can be obtained as a limit of the quantum commutator by
means of the Wigner transformation (see section ~\ref{sec:2}). Such
classical limit preserves a number of mathematical properties of the
original quantum commutator and this makes the classical dynamics
internally consistent. By internally consistent we mean that the
classical dynamics does not give any clue that it is just an
approximation since it displays all the correct properties that one
would expect (see below in this section). Let us now consider a
quantum system with two subsystems or sectors, which in general will
be mutually interacting. One can ask whether it is possible to take
the classical limit in just one of the sectors and still have an
internally consistent dynamics for the resulting mixed
quantum-classical system. After some definitions, this becomes a
mathematical problem with some physical input which will be called the
semiquantization problem here. As noted above, quantum-classical
mixed systems exists abundantly in the literature, where they are
understood as approximations to a fully quantum dynamics. They are not
meant to be consistent so they are not under debate here.

Let us clarify in what sense classical mechanics is a consistent limit
of the quantum mechanics and, in passing, introduce some notation. It
is a common feature of both classical and quantum mechanics that the
dynamics can be described in Heisenberg picture by an evolution
equation of the form
\begin{equation}
\frac{d A}{dt}= (A,H)+\frac{\partial A}{\partial t}\,,
\label{eq:1}
\end{equation}
where, $t$ is the time, $A$ is an arbitrary observable and $H$ is the
Hamiltonian of the system. The term $\frac{\partial A}{\partial t}$
takes into account the intrinsic time dependence of $A$. On the other
hand, the term $(A,H)$ describes the dynamic evolution of $A$. In the
quantum case the observables are self-adjoint operators in a Hilbert
space and the bracket $(\,,)$ is essentially the commutator. In the
classical case the observables are real functions on the phase space
and the dynamical bracket is the Poisson bracket:
\begin{eqnarray}
(A,B)_q &=& \frac{1}{i\hbar}[A,B]= \frac{1}{i\hbar}(AB-BA)\,,\nonumber \\
(A,B)_c &=& \{A,B\} =
\sum_i\left(
\frac{\partial A}{\partial x_i}\frac{\partial B}{\partial k_i}
-\frac{\partial A}{\partial k_i}\frac{\partial B}{\partial x_i}
\right)\,.
\label{eq:2}
\end{eqnarray}

Mathematically, the quantum and classical brackets have a number of
remarkable properties. First, they are universal in the sense that
they are independent of the particular dynamics. The latter is
specified by the Hamiltonian which in principle can be any observable
of the system. Second, they are Lie brackets, that is, they are
linear, antisymmetric and satisfy the Jacobi identity:
\begin{eqnarray}
 &&(A,B)=-(B,A)\,, \nonumber \\
 &&((A,B),C)+((B,C),A)+((C,A),B)=0\,.
\end{eqnarray}
The antisymmetry of the bracket ensures that time independent
Hamiltonians are conserved. The linearity guarantees that if $A(t)$
and $B(t)$ are two observables which only have dynamical evolution
(i.e., without intrinsic time dependence), and $a$ and $b$ are two
real constants, the observable $C_1(t)=aA(t)+bB(t)$ is also free of
intrinsic time dependence. Likewise, the Jacobi identity ensures that
the observable $C_2(t)=(A(t),B(t))$ also evolves dynamically only:
\begin{equation}
\frac{dC_2}{dt}=
(\frac{dA}{dt},B)+(A,\frac{dB}{dt})
=((A,H),B)+(A,(B,H))= ((A,B),H)= (C_2,H)\,.
\end{equation}
The third equality requires the Jacobi identity since $A$, $B$ and $H$
can be arbitrary. In particular, this property ensures the preservation
of the canonical relations among canonical variables. Third, for any
Hamiltonian, the dynamic evolution operator $(\ ,H)$ is a derivation,
that is, satisfies Leibniz's rule:
\begin{equation}
(AB,H)=(A,H)B+A(B,H)\,.
\end{equation}
Being a derivation guarantees that the product of observables is
consistent with time evolution, i.e., the observable $C_3(t)=A(t)B(t)$
is free of intrinsic time dependence if $A$ and $B$ are. In
particular, this ensures that the commutation relations among
canonical variables are preserved. (Note that commutation relations
here means the commutator, which may or may not coincide with the
dynamical Lie bracket.) Fourth, the brackets are such that
the reality or hermiticity conditions on the observables (in the
classical or quantum cases respectively) are preserved by time
evolution.

Giving up these properties would imply that some of the previous
constructions are not preserved by time evolution and this would
introduce an intrinsic time dependence in the dynamics. Note that this
is different from the question of whether the dynamics is conservative
or not; a non conservative Hamiltonian $H(t)$ introduces a privileged
origin of time in the dynamics but then the Hamiltonian
$H^\prime(t)=H(t-\tau)$ defines a dynamics which is precisely the same
as before except that the time is shifted by $\tau$, provided that the
dynamical bracket has all the properties noted above. In the absence
of any of these properties there would be a universal privileged time,
universal meaning independent of the particular Hamiltonian
~\cite{Salcedo:1996jr}. Three further important remarks are, first,
that the equivalence between Schr\"odinger and Heisenberg pictures can
only be proved if the bracket is a derivation, since it requires that
the product of observables be preserved by time evolution, second,
when the bracket is a derivation, c-number observables are
automatically free of dynamic evolution (since
$(1,H)=(1^2,H)=2(1,H)=0$, the second equality coming from Leibniz
rule), otherwise this property requires an independent postulate, and
third, the Lie bracket property is essential if one wants the
dynamical system to carry representations of symmetry groups of
transformations using the observables as infinitesimal generators. The
operator $(\,,H)$ is just a particular case corresponding to the group
of time translations, hence generalizing what we have already said to
other transformation groups, in the absence of the Lie bracket
property the dynamics would introduce intrinsic violations of
rotational invariance, etc.

In \cite{Salcedo:1996jr} a study of the semiquantization problem was carried out 
making two natural assumptions, first that the semiquantized theory
should enjoy all mathematical properties common to both quantum and
classical dynamics and second that when the two sectors are decoupled
they should evolve as if they were isolated, according to their usual
quantum or classical dynamics. More precisely it was required the
existence of a Heisenberg picture, a canonical structure plus the
condition that the product of two observables were preserved by the
time evolution. It was found that under these assumptions the only
consistent dynamics are either purely quantum or purely classical. It
was also found that removing the canonical structure condition allows
other dynamics but they are trivial in the sense that the classical
variables do not inherit fluctuations from their coupling to the
quantum sector, that is, there is no quantum backreaction on the
classical sector. This is the case of the semiclassical dynamics
commented above, where the classical variables are coupled to the
expectation values of the relevant quantum observables.

In the present work, the consistency of a universal semiquantization
is studied assuming only a canonical structure or assuming only
physical positivity of the resulting theory.  We consider the simplest
systems such as those described by position-momentum pairs or field
theories of real scalar fields. In section ~\ref{sec:2} we study some
existent proposals of the universal type to the semiquantization
problem and show that they fail to be consistent. In
section~\ref{sec:3}, we find that any universal canonical
semiquantization fails to fulfill some natural requirements when the
two sectors are decoupled. In section ~\ref{sec:4} it is found that
the requirement of physical positivity of the semiquantum theory
prevents the existence of quantum backreaction or even the coupling
among the quantum and classical sectors. Section \ref{sec:5}
summarizes our conclusions.

\section{The quantum-classical bracket}
\label{sec:2}

The general setting is as follows. There is one quantum sector and one
classical sector. We will consider only systems which are described by
conjugate canonical variables of the type position and momentum, that
is, Hilbert spaces of the form L$^2$(R$^n$) in the quantum case. The
observables are formed out of the classical canonical variables $x_i$,
$k_i$, $i=1,\dots,n_c$ and the quantum ones $q_i$, $p_i$,
$i=1,\dots,n_q$. Therefore, they are functions defined on the phase
space of the classical sector which take values on operators on the
Hilbert space of the quantum sector. The classical variables are
commuting numbers whereas $[q_i,q_j]=[p_i,p_j]=0$,
$[q_i,p_j]=i\hbar\delta_{ij}$, as usual. The standard proposal for the
quantum-classical bracket is
\begin{equation}
(A,B)_s=(A,B)_q+ \frac{1}{2}((A,B)_c-(B,A)_c)\,.
\label{eq:10}
\end{equation}
(The Poisson bracket $\{A,B\}$ is defined by eq.~(\ref{eq:2}) also
when $A$ and $B$ are non commuting quantities.) This dynamical bracket
has been proposed by various authors
~\cite{Boucher:1988ua,Aleksandrov:1981,Prezhdo:1997gs} starting from
different considerations. It should be noted, however, that
\cite{Boucher:1988ua} uses a Schr\"odinger picture and so this bracket
is used only to evolve the density matrix. This will be discussed
further in section~\ref{sec:4}. The bracket of
~\cite{Anderson:1995tn}, $(A,B)_s^\prime=(A,B)_q+ (A,B)_c$, is similar
except that it is not antisymmetric. In \cite{Prezhdo:1997gs} a Wigner
representation is chosen for the quantum operators.

This bracket can be obtained as follows. Let us start from a fully
quantum system with two sectors. The Hilbert space will be ${\cal
H}_q\otimes{\cal H}_c$ with ${\cal H}_{q,c}= {\rm L}^2({\rm
R}^{n_{q,c}})$. In order to take a classical limit later, let us apply
a Wigner transformation to the sector ${\cal H}_c$:
\begin{equation}
A(x,k;\hbar_c)= \int d^{n_c}y e^{-iy\cdot k/\hbar_c}\langle
x+\frac{1}{2}y|\hat{A}|x-\frac{1}{2}y\rangle \,.
\label{eq:5}
\end{equation}
Here $\hat{A}$ is the original operator on the full Hilbert space
${\cal H}_q\otimes{\cal H}_c$.  $|x\rangle$ is a basis state with
well-defined position in the space ${\cal H}_c$ only. Therefore, $A$
is an operator on ${\cal H}_q$ and a function on the phase space
spanned by $x_i$ and $k_i$, $i=1,\dots,n_c$. This transformation can
be inverted so that $\hat{A}$ can be recovered from $A$, thus $A$ is a
faithful representation of $\hat{A}$. The representation will depend
on the positive parameter $\hbar_c$ which is entirely arbitrary.

The Wigner transformation naturally defines a product among functions
on the phase space, namely (with obvious notation) $A*B$ is defined as
the Wigner representation of $\hat{A}\hat{B}$. Of course, if $A$, $B$
are regarded as $\hbar_c$-independent functions, the operation
represented by $*$ will depend on $\hbar_c$ explicitly.  The
commutator $[\hat{A},\hat{B}]$ is represented by $[A,B]_*=A*B-B*A$ and
so the fully quantum dynamical bracket is represented by
$\frac{1}{i\hbar}[A,B]_*$:
\begin{equation}
(\hat{A},\hat{B})_q=\frac{1}{i\hbar}[\hat{A},\hat{B}]\to
 \frac{1}{i\hbar}[A,B]_* \,.
\end{equation}
In order to obtain the dynamical bracket of the mixed
quantum-classical system, it remains to take the classical limit in
the sector ${\cal H}_c$. This can be done using the identity
\begin{equation}
e^{ix\cdot k/\hbar_c}=(2\pi\hbar_c)^{n_c}
e^{i\hbar_c\partial_x\cdot\partial_k}\delta(x)\delta(k)\,,
\end{equation}
which allows to express the product $*$ as
\begin{equation}
(A*B)(x,k)=
e^{\frac{1}{2}i\hbar_c(\partial^{(A)}_x\cdot\partial^{(B)}_k
-\partial^{(A)}_k\cdot\partial^{(B)}_x)}A(x,k)B(x,k)\,.
\label{eq:6}
\end{equation}
Here, $\partial^{(A)}_x$ means derivative of the $x$ dependence in $A$
only, etc. This formula is convenient to study the limit of small
$\hbar_c$. An expansion in powers of $\hbar_c$ gives
\begin{equation}
A*B= AB+\frac{i\hbar_c}{2}\{A,B\}+O({\hbar_c}^2)\,.
\label{eq:7}
\end{equation}
Therefore, the dynamical bracket takes the form
\begin{equation}
\frac{1}{i\hbar}[A,B]_*= \frac{1}{i\hbar}[A,B]+
\frac{1}{2}\frac{\hbar_c}{\hbar}\left(\{A,B\}-\{B,A\}\right)
+\frac{\hbar_c}{\hbar}O(\hbar_c)\,.
\label{eq:8}
\end{equation}
Taking now $\hbar_c=\hbar$ and neglecting terms of $O(\hbar)$ one
gets
\begin{equation}
\frac{1}{i\hbar}[A,B]+
\frac{1}{2}\left(\{A,B\}-\{B,A\}\right) \,,
\end{equation}
which is just the quantum-classical bracket $(\,,\,)_s$ defined in
eq.~(\ref{eq:10}). The idea would be that neglecting higher order
terms in $\hbar_c$ corresponds to the classical limit in the sector
${\cal H}_c$. In fact, when there is no quantum sector present,
$n_q=0$, $A$ and $B$ are commuting quantities and the limit
$\hbar=\hbar_c\to 0$ of $\frac{1}{i\hbar}[A,B]_*$ is well-defined and
gives the Poisson bracket. On the other hand, if there is no
classical sector, $n_c=0$, all terms containing $\hbar_c$ vanish
(cf. eq.~(\ref{eq:6})) and the prescription reproduces the usual
quantum commutator.

The construction of this bracket is somewhat tricky and in fact it
does not define a consistent coupling among the classical and quantum
sectors. As already noted, it is not a derivation (it does not
satisfies Leibniz rule). Even if one does not insist on this
requirement, the bracket $(\,,\,)_s$ does not define a canonical
structure because it fails to fulfill the Jacobi identity. This is
readily checked by taking three observables $A=qx$, $B=qpx$ and
$C=pk^2$, where $q$, $p$ are position and momentum variables of a one
dimensional quantum subsystem, and $x$, $k$ refer to the position and
momentum of the classical subsystem, also one dimensional. By direct
computation one finds
\begin{equation}
((A,B)_s,C)_s+((B,C)_s,A)_s+((C,A)_s,B)_s=-\frac{1}{2}(i\hbar)^2\,.
\end{equation}
(In order to show that the Jacobi identity is violated, it is
necessary to use at least two cubic operators. The identity is
preserved if all the operators involved are at most quadratic in $x$,
$k$, $q$ and $p$. Also, for the identity to fail at least two of the
operators should be of mixed quantum-classical type.)

Since the product $A*B$ is just a (faithful) representation of the
ordinary product of operators, it is associative and thus the
corresponding commutator $[A,B]_*$ satisfies the Jacobi identity for
any value of $\hbar_c$. The violation of the Jacobi identity in
$(\,,\,)_s$ comes from the truncation of the commutator at $O(\hbar)$
after taking $\hbar_c=\hbar$. Expanding the exact (untruncated)
commutator in powers of $\hbar_c$ one finds
\begin{equation}
[A,B]_*={\cal C}_0(A,B)
+\hbar_c{\cal C}_1(A,B)
+\hbar_c^2{\cal C}_2(A,B)+\cdots \,,
\end{equation}
The coefficients ${\cal C}_n$ are independent of $\hbar_c$ by
definition and can be computed using eq.~(\ref{eq:6}). In particular,
the coefficients ${\cal C}_0$ and ${\cal C}_1$ can be read off from
eq.~(\ref{eq:8}), being the commutator and the Poisson bracket
respectively. The Jacobi identity then yields a separate identity for
each power of $\hbar_c$
\begin{eqnarray}
0 &=& {\cal C}_0({\cal C}_0(A,B),C)+\hbox{c.p.} \nonumber
\\
0 &=& {\cal C}_0({\cal C}_1(A,B),C)
+{\cal C}_1({\cal C}_0(A,B),C)
+\hbox{c.p.} \\
0 &=& {\cal C}_0({\cal C}_2(A,B),C)
+{\cal C}_1({\cal C}_1(A,B),C)
+{\cal C}_2({\cal C}_0(A,B),C) 
+\hbox{c.p.} \nonumber
\end{eqnarray}
(Where c.p. stands for cyclic permutations of $A$, $B$, and $C$.)
When there is just one quantum sector, the only non vanishing
coefficient is ${\cal C}_0$ and the first equation yields the Jacobi
identity for the commutator. On the other hand, when there is just a
classical sector, ${\cal C}_0$ vanishes and the third equation yields
the Jacobi identity for the Poisson bracket (of commuting
quantities). If there are two sectors ${\cal C}_0$ no longer vanishes
and keeping only ${\cal C}_0+\hbar_c{\cal C}_1$, as in $(\,,\,)_s$,
violates the Jacobi identity at $O({\hbar_c}^2)$. As a rule, the
operation of keeping only the leading order in the expansion preserves
the Jacobi identity since this can be seen as a limit case. In the
purely quantum case the leading order is ${\cal C}_0$ and in the
purely classical case it is ${\cal C}_1$. However, in general, working
with the series truncated at $O(\hbar_c^n)$ preserves the Jacobi
identity only modulo $O(\hbar_c^n)$.

Contrary to the opinion expressed in Ref.~\cite{Prezhdo:1997gs}, we
think a dynamics leading to the bracket in eq.~(\ref{eq:10}) cannot be
consistent.  In our opinion, it is clear that any dynamics in a
Heisenberg picture necessarily defines a concrete dynamical bracket
and a consistent dynamics can only yield a consistent bracket.
Therefore, stating that the resulting dynamical bracket is ``naive''
or ``simplistic'' ~\cite{Prezhdo:1997gs} does not solve the
consistency problem. As follows from our previous discussion, the
bracket $[\,,\,]_*$ is indeed perfectly consistent, so the bracket
$(\,,\,)_s$ would enjoy the Jacobi identity if it were a true limit
case of $[\,,\,]_*$, but it is not. It is simply an arbitrary
prescription. This fact, as well as the fact that the Jacobi identity
is a non linear relation, accounts for the failure of that bracket to
be consistent.

This is a good place to illustrate what we mean by an internally
consistent semiquantization. The semiquantization given by $(\,,\,)_s$
is not internally consistent since checking the Lie bracket property
one finds that it is not preserved, therefore it can be concluded,
without using external information, that something is missing and that
$(\,,\,)_s$ is intrinsically an approximation. As discussed above, a
non-Lie dynamical bracket introduces violations of symmetries and can
have a limited range of applicability only.

Since $(\,,\,)_s$ violates Jacobi at second order in $\hbar_c$, it
would make sense to add the second order term ${\cal C}_2$, so that
Jacobi is preserved. However, new violations would appear at higher
orders and, in addition, the classical sector would be less
``classical''. It is to be expected that a systematic correction of
the bracket in order to exactly fulfill the Jacobi identity would end
up in a quantum-quantum system (except, may be if only particular
subclasses of observables are involved). At this point we can recall
that a similar conclusion was reached by De Witt long ago
~\cite{DeWitt:1962}. He showed that consistency with the uncertainty
principle in the quantum sector, requires to introduce systematic
corrections in the system in such a way that one ends up with a fully
quantum system in both sectors.  Of course, there is no problem in
having a mixed system if the two sectors are never coupled, but in
general, only fully classical or fully quantum dynamics are
consistent. His construction pursued to obtain a consistent action
functional to describe the mixed system thus effectively implying a
canonical structure.

\section{Obstructions to a canonical semiquantization}
\label{sec:3}

The semiquantization problem, that is, the construction of a
consistent dynamics for a mixed quantum-classical system, reminds the
quantization problem. The quantization problem in its most naive form
consists in associating to each function $A(x,k)$ on a classical phase
space an operator $\hat{A}=A(q,p)$ in L$^2$(R$^n$) using the
quantization rules $x_i\to q_i$, $k_i\to p_i$ and $1\to I$ (the
identity operator) and choosing the ordering of the operators in such
a way that $\{A,B\}\to (i\hbar)^{-1}[\hat{A},\hat{B}]$. As is known,
the quantization problem posed in this form does not have a
solution~\cite{VanHove:1951}; for arbitrary functions there is no way
to choose the order of the operators so that the Poisson bracket goes
into commutator. The semiquantization problem can also be seen as a
problem of ordering of operators since the trouble comes because
$\{A,B\}$ fails to fulfill the Jacobi identity when $A$ and $B$ are
not commuting. As we have seen, the naive antisymmetrization implied
by $(\,,\,)_s$ is insufficient to produce a Lie bracket in
general. The naive quantization problem has a solution when restricted
to the subspace of quadratic operators (namely, using $xk\to
\frac{1}{2}(qp+pq)$) and likewise $(\,,\,)_s$ is a Lie bracket when
restricted to the subspace of at most quadratic operators. This is
because in that case the coefficients ${\cal C}_n$ vanish for $n\ge
2$.

In order to solve the semiquantization problem, we could start trying
different combinations of commutators and Poisson brackets, or
equivalently different orderings for the operators. Instead of that,
we will show that under very general conditions the problem does not
have a solution within the canonical framework. This puts a strong
constraint on the kind of semiquantizations one should look for.

Let ${\cal A}_c$ be the set of classical observables, i.e., real
functions on the phase space of the classical sector, and ${\cal A}_q$
be set of quantum observables, operators on the Hilbert space of the
quantum sector. The full set of observables is ${\cal A}= {\cal
A}_c\otimes{\cal A}_q$, so a general observable will be of the form
$A=\sum_{ij}C_iQ_j$ where $C_i$ and $Q_j$ are purely classical and
purely quantum observables, respectively. Let $(\,,\,)$ denote the
semiquantum dynamical bracket, which will be assumed to be a Lie
bracket. We will consider dynamics satisfying the following postulates
\begin{equation}
(CQ,C^\prime) = (C,C^\prime)_cQ \,,\quad  
(CQ,Q^\prime) = (Q,Q^\prime)_qC \,,
\label{eq:17}
\end{equation}
for arbitrary purely classical observables $C$, $C^\prime$ and
arbitrary purely quantum observables $Q$, $Q^\prime$. These postulates
can be justified as follows. In a system formed by two quantum
subsystems, the observables of one sector commute those of the other
sector and so $[A_1A_2,A_1^\prime]= [A_1,A_1^\prime]A_2$ for arbitrary
$A_1$, $A_1^\prime$ in one sector and $A_2$ in the other sector. If
the quantum-classical system is a limit of the quantum-quantum one
obtains the above postulates. Note that the postulates are free of any
ordering problem. For another argument, consider that the Hamiltonian
of the semiquantum system is of the form $H= C^\prime+Q^\prime$. Since
both sectors are not coupled by an interaction term, each sector should
evolve separately as if it were isolated. This implies that when the
Hamiltonian is purely classical, a classical observable should evolve
classically and furthermore, a quantum observable must not
evolve. Reversing the roles of quantum and classical, and putting this
in infinitesimal form it follows that
\begin{equation}
(C,C^\prime) = (C,C^\prime)_c \,,\quad (Q,Q^\prime) =
(Q,Q^\prime)_q \,, \quad  (Q,C)=0\,.
\label{eq:18}
\end{equation}
These relations are weaker than our postulates. On the other hand, our
postulates can be derived from these ones if in addition it is assumed
that $(\,,C)$ and $(\,,Q)$ should be derivations. By themselves the
axioms in eqs.~(\ref{eq:18}) are too weak to sufficiently constrain
the form of the bracket. In order to be able to draw definite
conclusions we will make a stronger assumption which is equivalent to
our postulates in eqs.~(\ref{eq:17}) . Namely, we demand that when the
Hamiltonian is classical, an observable $QC$ should evolve into
$QC(t)$ where $C(t)$ is the classical evolution of $C$. In
infinitesimal form this yields the first postulate. The second
postulate follows similarly. This constraint on the form of the
evolution of $QC$ is automatically satisfied if the quantum-classical
system derives as a limit from a quantum-quantum system and therefore
it is a very natural requirement. Note that there is an implicit
assumption of universality in the argument, i.e. the dynamical bracket
should be the same for all Hamiltonians and any observable can be a
Hamiltonian. Then we can state the following theorem:

{\bf Theorem 1:} Let ${\cal A}={\cal A}_c\otimes{\cal A}_q$ be of the
position-momentum type in both sectors. Then, no Lie bracket $(\,,\,)$
in ${\cal A}$ can fulfill the axioms
\begin{equation}
(A,C) = (A,C)_c \,,\quad  
(A,Q) = (A,Q)_q \,
\label{eq:19}
\end{equation}
for all $C\in{\cal A}_c$, $Q\in{\cal A}_q$ and $A\in{\cal A}$.

Note that, because all observables are of the form $\sum_{ij}C_iQ_j$,
these axioms are equivalent to those in eq.~(\ref{eq:17}). The bracket
$(\,,\,)_s$ satisfies these axioms and thus it is a particular case.

It should be remarked that the theorem only applies to systems
described by position and momentum conjugate variables. Other
quantum-classical mixtures, e.g., a quantum sector with a finite
dimensional Hilbert space such as a spin system plus some classical
sector, are not directly ruled out by this theorem. Also, the
incompatibility refers to a semiquantization of the complete class of
observables. As noted above, the bracket $(\,,\,)_s$, which fulfills
our postulates, is a Lie bracket in the restricted subclass of
observables which are at most quadratic in $q,p,x,k$.

What follows in this section is devoted to the proof of this theorem.

In order to prove the incompatibility stated in the theorem, let us
assume that $(\,,\,)$ is a Lie bracket which satisfies our axioms. It
will be sufficient to consider a system with a one-dimensional quantum
sector and a one-dimensional classical sector. Also, in what follows
we will take $\hbar=1$ since keeping $\hbar$ variable (but strictly
positive) does not add anything to the proof. Let us consider the set
of observables
\begin{equation}
e_r=e^c_re^q_r\,,\quad e^c_r=e^{ik_rx-ix_rk}\,,\quad
e^q_r= e^{ip_rq-iq_rp}\,.
\end{equation}
Where $x_r$, $k_r$, $q_r$ and $p_r$ are arbitrary real numbers and
$x$, $k$, $q$ and $p$ are the dynamical variables. The observables of
the form $e^c_r$ form a basis of ${\cal A}_c$ and those of the form
$e^q_r$ form a basis of ${\cal A}_q$. This latter statement is more
clearly seen by using the form $e^q_r= e^{-\frac{1}{2}ip_rq_r}
e^{ip_rq}e^{-iq_rp}$ since $e^{ip_rq}$ and $e^{-iq_rp}$ are basis of
the operators which are functions of $q$ and $p$ respectively and any
operator in L$^2$(R) can be normal ordered putting the $q$ at the left
of the $p$. Therefore, $e_r$ defines a (linear) basis of ${\cal A}$.
Let us see that the bracket can be determined up to a c-number
function. Using the postulates, it is immediate that
\begin{equation}
(e_r,x)=ix_re_r\,,\quad 
(e_r,k)=ik_re_r\,,\quad
(e_r,q)=iq_re_r\,,\quad(e_r,p)=ip_re_r\,.
\end{equation}
The Jacobi identity can be expressed as
\begin{equation}
\delta_C(A,B)=(\delta_C A,B)+(A, \delta_C B)\,,\quad \delta_C :=(\,, C)\,,
\label{eq:22}
\end{equation}
for arbitrary $A$, $B$ and $C$. Therefore, the Jacobi identity requires
\begin{eqnarray}
((e_r,e_s),x) &=& i(x_r+x_s)(e_r,e_s)\,, \quad
((e_r,e_s),k)  =  i(k_r+k_s)(e_r,e_s)\,, \nonumber \\ 
((e_r,e_s),q) &=& i(q_r+q_s)(e_r,e_s)\,,  \quad
((e_r,e_s),p)  =  i(p_r+p_s)(e_r,e_s)\,.
\end{eqnarray}
On the other hand, $(e_r,e_s)$ will be a linear combination of the
$e_t$ and the previous equations imply that $x_t=x_r+x_s$,
$k_t=k_r+k_s$, $q_t=q_r+q_s$ and $p_t=p_r+p_s$. In summary,
\begin{eqnarray}
(e_r,e_s) &=& 
F(q_r,p_r,x_r,k_r;q_s,p_s,x_s,k_s)
e^{i(k_r+k_s)x-i(x_r+x_s)k+i(p_r+p_s)q-i(q_r+q_s)p} \nonumber \\
&:=& F_{rs}e_{r+s} \,.
\label{eq:21}
\end{eqnarray}
Here $F$ is some real function which depends on the particular bracket
only. The postulates are consistent with this form and correspond to
\begin{eqnarray}
F(q_r,p_r,x_r,k_r;0,0,x_s,k_s) &:=& F^c_{rs} =  v_{rs}\,,
\nonumber \\
F(q_r,p_r,x_r,k_r;q_s,p_s,0,0) &:=& F^q_{rs} = 2\sin(\frac{u_{rs}}{2})\,,
\label{eq:25}
\end{eqnarray}
where we have introduced the variables
\begin{equation}
u_{rs}=p_rq_s-q_rp_s\,,\quad v_{rs}= k_rx_s-x_rk_s\,.
\end{equation}
The functions $F^c$ and $F^q$ come from computing $(e^c_r,e^c_s)_c$
and $(e^q_r,e^q_s)_q$, respectively.

Up to now we have imposed the Jacobi identity only when one of the
operators is $x$, $k$, $q$ or $p$. The full Jacobi identity follows
from considering $((e_r,e_s),e_t)$. It is immediate that the Lie
bracket property can be expressed as
\begin{eqnarray}
 && F_{rs} = -F_{sr} \,,\quad \nonumber \\
 && F_{rs}F_{r+s,t}+F_{st}F_{s+t,r}+F_{tr}F_{t+r,s} = 0 \,.
\label{eq:26}
\end{eqnarray}
It is interesting to note that eqs.~(\ref{eq:26}) are valid in a
purely classical or purely quantum case. So $F^c$ and $F^q$ (and in
fact, their generalization for any number of degrees of freedom) are
solutions of those relations. The quantum-classical bracket of the
previous section gives
\begin{equation}
F^s_{rs}= 2\sin(\frac{u_{rs}}{2})
+ v_{rs}\cos(\frac{u_{rs}}{2})\,.
\end{equation}
This bracket comes from a classical expansion (but not a limit) around
the quantum-quantum case:
\begin{eqnarray}
F^{qq}_{rs} &=&
2\sin(\frac{u_{rs}}{2})\cos(\frac{v_{rs}}{2})
+ 2\cos(\frac{u_{rs}}{2})\sin(\frac{v_{rs}}{2}) \nonumber \\
&=&
2\sin(\frac{u_{rs}}{2} + \frac{v_{rs}}{2})\,,
\end{eqnarray}
(which, of course, is equivalent to a two dimensional quantum case).
$F^s$ satisfies the axioms, eqs.~(\ref{eq:25}), but fails to fulfill
the Jacobi identity, second eq.~(\ref{eq:26}). Let us show that the
eqs.~(\ref{eq:25}) and (\ref{eq:26}) are, in fact, incompatible. As
proven in appendix~\ref{app:A}, these equations imply that $F$ can
only depend on the combinations $u$ and $v$ introduced above, that is
\begin{equation}
F_{rs}=F(u_{rs},v_{rs})\,.
\label{EQ:29}
\end{equation}
This follows only from the Lie bracket property of $F$ and the fact
that the boundary conditions, eqs.~(\ref{eq:25}), depend also on $u$
and $v$. Using eq.~(\ref{EQ:29}), the postulates become
\begin{equation}
F(u,0)= 2\sin(u/2) \,, \quad  F(0,v)= v\,,
\label{eq:31}
\end{equation}
and the Lie bracket conditions become
\begin{eqnarray}
&& F(u,v)=-F(-u,-v)\,, \nonumber \\ &&
F(u_{rs},v_{rs})F(u_{rt}+u_{st},v_{rt}+v_{st})+\hbox{c.p.} =0 \,.
\label{eq:30}
\end{eqnarray}
As proven in appendix~\ref{app:A}, these Lie bracket conditions plus
$\partial_uF(0,0)=\partial_vF(0,0)=1$, only admit the solutions
\begin{equation}
F(u,v)=\frac{1}{h}\sin(hu+hv)\,,
\label{EQ:33}
\end{equation}
($h$ being and arbitrary constant) or the degenerated case 
$F(u,v)=u+v$, which cannot accommodate the
two postulates, eqs.~(\ref{eq:31}), for any value of $h$.
This implies that the function $F$ is only consistent with Jacobi if
it is purely quantum $F(x)=2\sin(x/2)$ or purely classical $F(x)=x$,
but does not admit mixed types. This completes the proof.

The proof of the incompatibility of the axioms can also be done by an
alternative method which is constructive (but requires to perform
symbolic calculations with the help of a computer). We will devote the
reminder of this section to discuss this method.

Let us consider a basis of the space ${\cal A}$ with observables of
the form $q^rp^sx^tk^\ell$, where $r,s,t,\ell= 0,1,\dots$. (The order
$qpxk$ will be taken as the canonical order of these variables.) Each
basis element can be assigned a degree given by $n=r+s+t+\ell$. Let us
use the notation $C_n$, $Q_n$, $M_n$ and $A_n$ to refer to basis
elements of degree $n$ which are a purely classical (i.e., $r=s=0$),
purely quantum ($t=\ell=0$), mixed quantum-classical ($r+s > 0$ and
$t+\ell > 0$) and arbitrary, respectively. Using the axioms, the
brackets of each pair of basis elements can be worked out except those
of the form $(M_n,M_{n^\prime})$ (where necessarily $n,n^\prime \ge
2$). Using the Jacobi identity, as in eq.~(\ref{eq:22}), for
$\xi=q,p,x,k$, yields
\begin{eqnarray}
((M_n,M_{n^\prime}),\xi) &=& 
((M_n,\xi),M_{n^\prime})+ (M_n,(M_{n^\prime},\xi))
\nonumber \\
&=& (M_{n-1},M_{n^\prime})+ (M_n,M_{n^\prime-1})\,.
\end{eqnarray}
Since the knowledge of $(A,\xi)$ for $\xi=q,p,x,k$ determines $A$
completely up to an additive c-number constant, this relation allows
to determine $(M_n,M_{n^\prime})$ by induction up to an additive
constant. Such constants play a similar role as the function $F$ in
the plain wave basis used above. Their number increases rapidly with
$n$ and $n^\prime$.

We proceed by selecting values for these constants so that the Jacobi
identity is fulfilled for arbitrary operators, if possible. Because
the Jacobi identity is trivial unless two of the operators involved
are of mixed type, only the case $\langle M,M,A\rangle$ gives
information on the constants. At step 1 we consider the brackets of
the form $(M_2,M_2)$ which contain 6 unknown constants. These constants
are uniquely determined imposing the Jacobi identity to the triples
$\langle M_2,M_2,Q_2 \rangle$ and $\langle M_2,M_2,C_2\rangle$. The
Jacobi identity for $\langle M_2,M_2,M_2\rangle$ turns out to be
fulfilled automatically. From now on, these 6 constants are fixed to
their unique value. At step 2, we consider $(M_2,M_3)$ which contain
48 unknowns. These constants are uniquely determined imposing Jacobi
to $\langle M_2,M_3,Q_2\rangle$ and $\langle M_2,M_3,C_2\rangle$, and
again Jacobi for $\langle M_2,A_3,M_2 \rangle$ comes out
automatically. The 48 unknowns are then fixed to their unique value.
At step 3, the 100 unknowns in $(M_2,M_4)$ are fixed to their unique
value which is determined from $\langle M_2,M_4,Q_2\rangle$ and
$\langle M_2,M_4,C_2\rangle$, and Jacobi for $\langle
M_2,A_4,M_2\rangle$ is automatic. All these constants are those
corresponding to the bracket $(\,,\,)_s$ of section~\ref{sec:2}.
Finally, at step 4 the procedure breaks down. The 66 unknowns of
$(M_3,M_3)$ are uniquely determined by $\langle M_3,Q_3,M_2\rangle$
and $\langle M_3,C_3,M_2\rangle$, but they turn out to be inconsistent
with the Jacobi identity for $\langle M_3,M_3,M_2\rangle$.

\section{Quantum backreaction and positivity}
\label{sec:4}

In this section we want to discuss obstructions to mixing quantum and
classical systems not related to a canonical structure but to the
requirement of positivity.

Here we no longer require the existence of a canonical structure plus
a Heisenberg picture, etc. In view of this we will need a definition
of what it is meant by a quantum-classical system. We will demand that
the quantum variables obey to the usual quantum commutation relations
and the classical variables commute. Because the quantum variables do
not commute, the usual proof shows that they must be subjected to the
uncertainty principle. They have primary quantum fluctuations. (They
are primary because they exist even in the absence of a coupling to
other degrees of freedom.) If the two sectors are coupled, the
classical observables may have induced fluctuations, the so-called
quantum backreaction. The two requirements of primary fluctuations in
the quantum variables and commutation of the classical variables is
just what we mean by ``quantum-classical'' mixing. We want to argue
that even these weak conditions, plus another natural requirement,
namely, that the mixed quantum-classical system must correspond to
some limit case of a quantum-quantum system, forbid the possibility of
a quantum backreaction on the classical sector. Since such secondary
fluctuations are expected to appear whenever the two sectors are
coupled by an interaction term, this would suggest that no consistent
quantum-classical mixing exists.

The requirement that the mixed system should be a limit of a full
quantum system seems weak but still has non trivial consequences: if
some quantity of the quantum-quantum system is always positive, the
corresponding quantity must at least be non negative in the
quantum-classical system. For instance, in the purely quantum theory
the variance of any observable must be non negative, i.e., $\langle
A^2\rangle \ge \langle A\rangle^2$, where $\langle \,\rangle$ refers
to the quantum average. The equal sign corresponds to an observable
which is free of quantum fluctuations and so it implies $\langle
f(A)\rangle = f(\langle A\rangle)$ for any function $f(x)$ as
well. Then the same properties must hold in the mixed
quantum-classical system. Of course, they hold in the purely classical
system in particular. This is equivalent to say that whatever is the
measure corresponding to the fluctuations in the mixed system (induced
by the quantum sector) it must be non negative.

As is well-known in quantum mechanics, if the commutator of two
observables does not vanish on some quantum state vector, at least one
of the observables must present quantum fluctuations in that state.
Here we will need a reciprocal of this statement:

{\bf Theorem 2:} In a purely quantum theory, let $|\psi_0\rangle$ be the
normalized ground state and $A$ any observable, then i)
\begin{equation}
0 \le
-\frac{i\hbar}{2}\langle[A,\dot A]\rangle_{\psi_0}\,,
\label{eq:35}
\end{equation}
and ii) when the ground state is not degenerated, the equal sign holds
if and only if $A$ is free of dispersion in $|\psi_0\rangle$.

This can be proved as follows. Let $H$ be the Hamiltonian and $E_0$
the ground state energy, $H|\psi_0\rangle=E_0|\psi_0\rangle$, then,
because $H-E_0$ is non negative and $A$ is self-adjoint, the operator
$A(H-E_0)A$ is also non negative. Therefore
\begin{equation}
0\le \langle A(H-E_0)A\rangle_{\psi_0}=
-\frac{1}{2}\langle[A,[A,H]]\rangle_{\psi_0}\,.
\end{equation}
Eq.~(\ref{eq:35}) follows then from the relation $i\hbar\dot A=
[A,H]$. To show ii), assume that $A$ has no dispersion in the ground
state, $\langle A^2\rangle_{\psi_0}=\langle A\rangle^2_{\psi_0}$, then
$A|\psi_0\rangle$ is proportional to $|\psi_0\rangle$ and
automatically $\langle[A,B]\rangle_{\psi_0}=0$ for any
$B$. Conversely, assume that the equal sign applies in
eq.~(\ref{eq:35}), then $0=\langle A(H-E_0)A\rangle_{\psi_0}$, and so
$A|\psi_0\rangle$ is another ground state. If this is not degenerated
$A|\psi_0\rangle$ must be proportional to $|\psi_0\rangle$ and $A$ is
free of dispersion in the ground state.

A corollary of this theorem is that when $A$ commutes with $\dot A$,
$A$ cannot have quantum fluctuations in a non degenerated ground
state, so in some sense it is a reciprocal of the argument leading to
the uncertainty principle. It has immediate consequences to the
semiquantization problem. For a mixed system at zero temperature (and
so in the ground state) the classical variables $x_i$ and
$k_i=\dot{x}_i$ will commute with each other and, being the limit of a
quantum-quantum system, they will be free of quantum fluctuations. In
other words, since any induced quantum fluctuations would spoil the
commutativity of the classical variables, in a quantum-classical
mixing there cannot be quantum backreaction on the classical sector.

It should be noted that there are actually proposals of mixed
quantum-classical systems in which the classical variables commute and
at the same time have secondary quantum fluctuations. Such
prescriptions exists, without invalidating our conclusions above,
because they violate positivity of the measure of the quantum
fluctuations. That is, positive observables do not have a positive
expectation value. A first example is the proposal in
~\cite{Boucher:1988ua}. There the quantum-classical system is
described in terms of a density matrix which depends on the quantum
variables $q,p$ and the classical variables $x,k$. The evolution is
described in the Schr\"odinger picture in the form $\dot\rho=(\rho,H)$
where $H$ is the Hamiltonian and $(\,,\,)$ is the dynamical bracket.
In ~\cite{Boucher:1988ua} the bracket is completely determined by
imposing several natural requirements; it should reduce to the
commutator or Poisson bracket as particular cases, the evolution
preserves the hermiticity and the trace of the density matrix, and it
is invariant under classical canonical transformations and quantum
unitary transformations. The result is again the bracket $(\,,\,)_s$
in eq.~(\ref{eq:10}). As noted by the authors, there is, however, one
essential requirement which is violated by this construction, namely,
if one starts with a positive density matrix $\rho$, its positivity is
not preserved by the evolution, in general. It is noteworthy that,
since the observables do not evolve, this construction does not
introduce any intrinsic time dependence in the dynamics. For instance,
the commutator $[q,p]=i\hbar$ is automatically time independent and
the classical variables always commute. As noted, when $(\,,H)$ is not
a derivation, the Heisenberg and Schr\"odinger dynamics are no longer
equivalent and there is no contradiction with our discussion above,
which refers to the Heisenberg picture.  Nevertheless, because Jacobi
is not satisfied, there will be problems implementing time-independent
symmetry transformations, in addition to the positivity problem.

Another proposal is that of~\cite{Salcedo:1994sn}. There, it is noted
that the stochastic quantization
program~\cite{Parisi:1981ys,Damgaard:1987rr,Namiki:1992} leads to a
natural definition of a semiquantized dynamics. As it may be recalled,
in the stochastic quantization approach the dynamical variables evolve
in a fictitious time, the simulation time, following a stochastic
differential equation, the Langevin equation, which corresponds to a
particular Monte Carlo method to sample the Euclidean path integral of
the system. For instance, let $S[\phi]$ be the Euclidean action of
quantum field theory with $n$ scalar fields $\phi_i(x)$, $i=1,\dots,n$
in a flat four dimensional space-time. Then the functional integral
with Boltzmann weight $e^{-S/\hbar}$ is correctly sampled by the
equilibrium distribution of a random walk described by the following
Langevin equation ~\cite{Damgaard:1987rr}
\begin{equation}
\frac{\partial\phi_i(x;\tau)}{\partial\tau}
=-\frac{\delta S[\phi]}
{\delta\phi_i(x;\tau)}
+\sqrt{\hbar}\eta_i(x;\tau) \,.
\end{equation}
Here $\tau$ is the simulation time and $\eta_i(x;\tau)$ are
independent stochastic centered Gaussian variables normalized to
\begin{equation}
\langle\!\langle\eta_i(x;\tau)\eta_{i'}(x';\tau')\rangle\!\rangle =
2\delta_{ii'}\delta(x-x')\delta(\tau-\tau')\,.
\end{equation}
The variables $\eta_i(x;\tau)$ introduce the quantum fluctuations in
the system. In their absence, the fields $\phi$ would fall into a
solution of the (Euclidean) classical equations of motion, $\delta
S/\delta\phi_i(x)=0$.

The $\hbar$ dependence of the Langevin equation suggests a natural
definition for the semiquantized system~\cite{Salcedo:1994sn}, namely, to
replace $\hbar$ by $\hbar_i=0,1$ where $1$ corresponds to a quantum
degree of freedom and $0$ to a classical one. (We will use units
$\hbar=1$ from now on.) The classical degrees of freedom will not have
primary quantum fluctuations but, if they are coupled to the quantum
sector, they will present induced secondary fluctuations. In order to
see what consequences follow from such proposal, let us take the
example studied in \cite{Salcedo:1994sn}. Consider a system composed by two
relativistic fields with a quadratic action
\begin{equation}
S(\phi_1,\phi_2)=\int d^4x\left(
\frac{1}{2}(\partial\phi_1)^2+\frac{1}{2}m_1^2\phi_1^2+
\frac{1}{2}(\partial\phi_2)^2+\frac{1}{2}m_1^2\phi_2^2+g\phi_1\phi_2
\right)\,.
\end{equation}
Since the action is translationally invariant, it is convenient to use
a momentum representation:
\begin{equation}
S(\phi_1,\phi_2)=\int \frac{d^4k}{(2\pi)^4}
\,\frac{1}{2}\Phi^\dagger(k) M(k)\Phi(k)\,,
\end{equation}
where
\begin{equation}
\quad \Phi(k)=\pmatrix{\tilde\phi_1(k) \cr \tilde\phi_2(k)} \,,
\quad M(k)=\pmatrix{k^2+m_1^2 & g \cr g & k ^2+m_2^2 \cr}\,
\end{equation}
and $\tilde\phi_i(k)$ is the Fourier transform of $\phi_i(x)$. As
usual, we will assume $m_1^2,m_2^2>0$ and $m_1^2m_2^2>g^2$, so that
$M(k)$ is positive definite. Because the action is quadratic, the
equilibrium solution of the Langevin equation can be solved in closed
form. The connected two-point function or propagator is given in
momentum space by the following matrix~\cite{Salcedo:1994sn}
\begin{equation}
W(k)= \frac{\hbar_2(k^2+m_1^2) +\hbar_1(k^2+m_2^2)}
{(k^2+m_1^2) +(k^2+m_2^2)}W_Q(k) + \frac{\hbar_1-\hbar_2}
{(k^2+m_1^2) +(k^2+m_2^2)}\sigma_z\,,
\end{equation}
where $\sigma_z$ refers to the Pauli matrix ${\rm diag}(1,-1)$, and
$W_Q(k)$ is the inverse matrix of $M(k)$. In this formula the
$\hbar_i$ are arbitrary non negative numbers. The propagator in
$x$-space is
\begin{equation}
\langle T\phi_i(y)\phi_j(x)\rangle = \int\frac{d^4k}{(2\pi)^4}
e^{-ik(y-x)}\,W_{ij}(k) \,.
\end{equation}
This Green's function is directly connected since $\langle
\phi_i(x)\rangle=0$. It can be shown~\cite{Salcedo:1994sn} that there are no
connected Green's functions of three or more points, so the system is
Gaussian.

From the form of $W(k)$ it follows that in the fully classical case,
$\hbar_i=0$, the connected two-point function vanishes implying that
the fields are free from fluctuations. On the other hand, in the fully
quantum case, $\hbar_i=1$, $W$ is just $W_Q=M^{-1}$ which is the
standard quantum propagator. If $\hbar_1=1$ and $\hbar_2=0$, and, in
addition, $g\ne 0$, $\langle (\phi_2(x))^2\rangle$ will not vanish and
thus $\phi_2$ is subjected to induced secondary fluctuations. When
$g=0$ both sectors are decoupled.

On the other hand, we can obtain the equal-time commutation relations
of the fields by considering the large momentum limit of the propagator:
\begin{equation}
W(k) = \frac{1}{k^2}
\pmatrix{ \hbar_1 & 0 \cr 0 & \hbar_2 \cr } + O(\frac{1}{k^4})\,.
\label{eq:44}
\end{equation}
This directly implies
\begin{equation} 
\delta(y^0-x^0)\langle [\phi_i(y),\dot\phi_j(x)]\rangle =
\hbar_i\delta_{ij}\delta(x-y)\,.
\label{eq:45}
\end{equation}
Therefore, if $\hbar_1=1$ and $\hbar_2=0$, the field $\phi_2$ will be
classical, in the sense that it commutes with its conjugate momentum,
even if it is subjected to quantum backreaction from the quantum field
$\phi_1$.

We have checked that this system is a quantum-classical mixture
according to our previous definition and also that there is quantum
backreaction, but we still have to see if it preserves positivity. It
can be shown~\cite{Salcedo:1994sn} that the matrix $W(k)$ is definite
positive for all momenta and arbitrary non negative
$\hbar_i$. However, as is well known, physical positivity corresponds
rather to the stronger requirement of reflection positivity in
Euclidean space~\cite{Glimm:1987}. Since the theory is quadratic, it
is sufficient to study the Lehmann representation of the propagator,
which comes from inserting a complete set of eigenstates:
\begin{equation}
W_{ij}(k)=
\int\,d\mu\frac{\rho_{ij}(\mu)}{k^2+\mu}
\end{equation}
where the spectral density is defined as
\begin{equation}
\rho_{ij}(q^2) = (2\pi)^3\sum_n\delta^4(p_n-q)
\langle 0|\phi_i(0)|n\rangle \langle n|\phi_j(0)|0\rangle \,.
\end{equation}
Reflection positivity requires $\rho(\mu)\ge 0$. For the purely
quantum case we have
\begin{equation}
W_Q(k) = \frac{P_+}{k^2+m_+^2} + \frac{P_-}{k^2+m_-^2} \,,
\end{equation}
where $P_\pm$ are the two orthogonal projectors onto the normal modes,
corresponding to diagonalize $M(k)$, and $m^2_\pm={1\over
2}(m_1^2+m_2^2 \pm R)$ are their squared masses (with $R =
\sqrt{(m_1^2-m_2^2)^2 +4g^2}$). For arbitrary $\hbar_i$ it is found
\begin{equation}
W(k) = \frac{Q_+}{k^2+m_+^2} + \frac{Q_-}{k^2+m_-^2} 
+ \frac{Q_3}{k^2 + m_3^2}
\label{eq:49}
\end{equation}
where $m_3^2=\frac{1}{2}(m_1^2+m_2^2)$, and
\begin{eqnarray}
Q_\pm &=& \left(
\frac{\hbar_1+\hbar_2}{2}\pm\frac{(\hbar_1-\hbar_2)(m_1^2-m_2^2)}{2R}
\right)P_\pm \,, 
\nonumber \\
Q_3 &=& \frac{\hbar_1-\hbar_2}{2}
\left(\sigma_z -\frac{m_1^2-m_2^2}{R}(P_+-P_-)\right) \,.
\end{eqnarray}
One can see that there is an extra mode, namely, $m_3^2$.
Unfortunately, whereas $Q_\pm$ are non negative, $Q_3$ is not in
general, since ${\tr}\,(Q_3) = 0$. This means that the covariance
matrix $W(k)$ is positive but not reflection positive except in the
trivial cases $\hbar_1=\hbar_2$ or $g=0$.  The latter case describes
two non interacting sectors, and the first case corresponds to two
classical sectors if $\hbar_1=\hbar_2=0$ or two quantum sectors if
$\hbar_1=\hbar_2>0$. Beyond these trivial cases, this theory does not
define a Hilbert space with positive definite metric, i.e., it does
not define a positive physical measure, and for instance, one can
construct operators with negative variance. In other words, the
probabilistic interpretation (of which the classical case is a limit)
breaks down. The theory must be rejected (or else work with a
restricted set of observables, which in this context would be ad hoc).

Mathematically, the lack of reflection positivity is a direct
consequence of the commutation relations eq.~(\ref{eq:45}). In effect,
as noted, the commutation relations are equivalent to
eq.~(\ref{eq:44}), and comparing with eq.~(\ref{eq:49}) for large
$k^2$, it follows that $\hbar_2= (Q_+ + Q_- + Q_3)_{22}$. Therefore,
if $\phi_2(x)$ is classical and so $\hbar_2=0$, the cancellation
requires $(Q_3)_{22}$ to be negative. This argument can be expected to
hold on general grounds. Indeed, we have the following theorem:

{\bf Theorem 3:} A theory of relativistic scalar fields which is quadratic,
reflection positive, translationally invariant and
$\langle\phi_i\rangle=0$, cannot have quantum and classical sectors
unless they are decoupled.

This can be proved as follows. Under the assumptions, all information
on the theory is contained in the propagator or equivalently in the
spectral density $\rho(\mu)$. For simplicity we consider theories with
one quantum field $\phi_1$ and one classical field $\phi_2$.  Being a
classical field means that $\phi_2$ and $\dot\phi_2$ commute at equal
time. Then $\langle T\phi_2(x)\phi_2(y)\rangle$ and $\langle
T\phi_2(x)\dot\phi_2(y)\rangle$ are continuous regarded as functions
of $t=x_0-y_0$ at $t=0$. This implies that the function $\langle
T\phi_2(x)\phi_2(y)\rangle$ is continuous and with continuous first
derivative at $t=0$. As a consequence its Fourier transform must be
$W_{22}(k)=O(\frac{1}{k^4})$ for large $k^2$. From the Lehmann
representation, this implies that $\rho_{22}(\mu)$ must average to
zero (otherwise $W_{22}(k)=O(\frac{1}{k^2})$) and then reflection
positivity requires $\rho_{22}=0$ everywhere. At this point we have
already shown that the connected propagator of a classical field must
vanish and so the classical field cannot have secondary quantum
fluctuations. This is in agreement with our Theorem 2 above. The
stronger statement in Theorem 3 comes from noting that if
$\rho_{22}=0$, positivity of the matrix $\rho$ requires
$\rho_{12}=\rho_{21}=0$ as well, therefore there is no mixing among
the two sectors.

\section{Summary and conclusions}
\label{sec:5}

In the present work, we study the internal consistency of
semiquantization schemes of the universal type. In the Introduction it
was argued that the classical dynamics is an internally consistent
limit of the quantum dynamics, since the Poisson bracket preserves a
number of essential properties of the quantum commutator (Lie bracket
property, Leibniz's rule and hermiticity). Both dynamics, quantum and
classical, are of the universal type since they have a fixed dynamical
bracket independent of the particular Hamiltonian. There it is also
discussed what unacceptable consequences would follow by giving up any
of the above mentioned properties, namely, intrinsic breaking of
symmetries and lack of hermiticity.

In section \ref{sec:2} the standard quantum-classical dynamical
bracket for a system with two sectors (each of the type
position-momentum) is derived as a ``limit'' of the quantum-quantum
bracket with the help of the Wigner representation. It is argued that
such a bracket is not internally consistent, since it fails to satisfy
both the Jacobi identity and the Leibniz's rule. It is pointed out
that this failure is due to the fact that such a bracket is not a true
limit case but rather a truncation at second order in an expansion in
$\hbar$. On the contrary the classical limit (in all sectors) is a
true limit since it keeps only the leading order in $\hbar$ and so it
preserves both the Jacobi identity and the Leibniz's rule.

In section \ref{sec:3} the semiquantization problem in its canonical
version is studied. It is pointed out the similarity of this problem
with that of the naive quantization of classical systems, which is
known not to have a solution for arbitrarily large spaces of
observables. Both problems can be tied to the ordering problem of
operators.

Roughly speaking, the canonical semiquantization problem consists in
finding a Lie bracket in the algebra of observables of the mixed
quantum-classical theory which interpolates between the Poisson
bracket and the quantum commutator. (The Leibniz's rule, which is
common to both classical and quantum dynamics, is not imposed on the
semiquantum bracket, since it was already shown in
\cite{Salcedo:1996jr} that such a requirement plus the Lie bracket
condition only allows for purely classical or purely quantum dynamics,
at least for systems of the position-momentum type.) However, in order
to determine the bracket of observables of mixed type, some
assumptions have to be made. Our assumptions are cast in the axioms in
eq.~(\ref{eq:17}). They follow either from assuming that the
quantum-classical system is a limit of a quantum-quantum system or
else from natural requirements on the behavior of the two sectors when
they are decoupled. Such axioms are sufficiently general as to cover
the case of the standard quantum-classical bracket of section
\ref{sec:2}. Theorem 1 is the main result of section \ref{sec:3}. It
states that, if both sectors are of the position-momentum type (and
are non trivial, i.e., they are not zero dimensional) no Lie bracket
exists which satisfies the axioms. In other words, under the
assumptions, there is no consistent canonical semiquatization. It
would be interesting to know whether the axioms allow for consistent
semiquantizations when one or both sectors are not of the
position-momentum type. For instance, one can consider a particle with
classical position and momentum but quantum spin. It would also be
interesting to know whether Theorem 1 can be adapted when the axioms
are weakened to those in eq.~(\ref{eq:18}), or else, to find what kind
of consistent semiquantizations are obtained.

In section \ref{sec:4} the requirement of a canonical structure for
the semiquantized system is dropped. It is only imposed that the
variables in the quantum sector are not commuting and those in the
classical sector are commuting, and further that positive observables
must have positive expectation values. This latter requirement follows
immediately if the mixed classical-quantum system is the limit of a
quantum-quantum system.

Being non commuting, the standard argument shows that the quantum
variables must have primary quantum fluctuations.  In section
\ref{sec:4} a sort of reciprocal of the uncertainty principle is
proven, Theorem 2, which implies that commuting variables cannot have
fluctuations. Therefore, under the assumptions, any semiquantization
scheme either does not have quantum backreaction on the classical
sector or else breaks positivity. One example of the first possibility
is semiclassical gravity. Two examples of the second possibility are
discussed in section \ref{sec:4}. Finally a further Theorem 3 is
presented which shows that, for scalar fields with
quadratic Lagrangians, positivity of the theory not only forbids any
quantum backreaction on the classical sector but also any coupling
among the two sectors.

These negative results, contained in the three theorems proven in the
text, put constraints to the form of possible semiquantization
approaches, especially those of universal type. A further negative
result has already been noted: quantum-classical field theories fail
to be renormalizable even if their corresponding quantum-quantum
version is renormalizable~\cite{Salcedo:1994sn}. Nevertheless, it
should be kept in mind that our conclusions only apply under the
assumptions made, and so more general forms of mixed quantum-classical
systems cannot be ruled out.

\section*{Acknowledgments}
This work is supported in part by funds provided by the Spanish DGICYT
grant no. PB95-1204 and Junta de Andaluc\'{\i}a grant no. FQM0225.

\appendix{}\section{Proof of eqs.~(\ref{EQ:29}) and ~(\ref{EQ:33})}
\label{app:A}

In order to prove eq.~(\ref{EQ:29}), let us consider the Jacobi
identity, eq.~(\ref{eq:26}), with $x_t=k_t=0$. This yields
\begin{equation}
F_{rs}F^q_{r+s,t}+F^q_{st}F_{t+s,r}+F^q_{tr}F_{t+r,s}=0 \,.
\end{equation}
Let us recall that $F^q_{rs}=2\sin(u_{rs}/2)$, with
$u_{rs}=p_rq_s-q_rp_s$.
Using the antisymmetry of $F$, the relation can be written as
\begin{equation}
F^q_{st}F_{r,s+t}=F^q_{r+s,t}F_{rs}-F^q_{rt}F_{r+t,s} \,.
\label{eq:A1}
\end{equation}
Note that $F_{rs}$, $F_{r,s+t}$ and $F_{r+t,s}$ all have the same
dependence on the classical variables $x_r,k_r,x_s,k_s$ and so these
variables can be treated just as fixed parameters in $F$, thus we can
concentrate on the dependence on the quantum sector only and use the
notation $F_{rs}=F(q_r,p_r;q_s,p_s)$. Now let us take $q_t=\lambda
q_r$ and $p_t=\lambda p_r$, then $u_{tr}=0$ and $u_{r+s,t}=u_{st}$ and
so $F^q_{tr}=0$ and $F^q_{r+s,t}=F^q_{st}$. Thus eq.~(\ref{eq:A1})
implies
\begin{equation}
F_{r,s+t}=F_{rs}\,,
\end{equation}
or in other words, the quantity $F(q_r,p_r;q_s+\lambda q_r,p_s+\lambda
p_r)$ is actually independent of $\lambda$. Choosing
$\lambda=-p_s/p_r$ yields
\begin{equation}
F_{rs}=F(q_r,p_r;\frac{u_{rs}}{p_r},0)\,.
\end{equation}
Likewise, using an analogous argument or using antisymmetry,
$F(q_r+\lambda q_s,p_r+\lambda p_s;q_s,p_s)$ is also independent of
$\lambda$, thus
\begin{equation}
F_{rs}=F(0,p_r;\frac{u_{rs}}{p_r},0)\,.
\end{equation}
This implies that $F_{rs}$ depends at most on $u_{rs}$ and $p_r$. A
similar argument shows that $F_{rs}$ may depend at most on $u_{rs}$
and $q_r$, therefore it depends only on $u_{rs}$.

Everything can be repeated for the dependence of $F$ on the classical
variables using that $F^c_{rs}=v_{rs}$ is an antisymmetric function of
$v_{rs}= k_rx_s-x_rk_s$. Thus, $F_{rs}$ depends only on $u_{rs}$ and
$v_{rs}$.

Next, let us prove eq.~(\ref{EQ:33}). First of all, note that the
three variables $x^u=u_{st}$, $y^u=u_{tr}$ and $z^u=u_{rs}$ are
independent, i.e, the triple $(x^u,y^u,z^u)$ can take any value in
R$^3$, and similarly $x^v$, $y^v$ and $z^v$ in the classical
sector. Second, the structure of eqs.~(\ref{eq:30}) is two dimensional
but can naturally be extended to any number of dimensions. Let us
denote by $x^i$, $i=1,\dots,n$ the corresponding variables, i.e.,
$F=F(x)$ (in our particular case, $n=2$ with $x^1=u$ and
$x^2=v$). Then the equations take the form
\begin{eqnarray}
&& F(x)=-F(-x)\,, \nonumber \\
&& F(x)F(y-z)+F(y)F(z-x)+F(z)F(x-y) =0 \,, \label{eq:A5}
\\
&& F_i(0)=1\,, \nonumber
\end{eqnarray}
where $F_i(x) := \partial_iF(x)$. The last relation comes from
$\partial_uF(0,0)=\partial_vF(0,0)=1$ which follows from the
postulates. (Actually, only $\partial_iF(0)\ne 0$ is essential for the
present argument). It is clear from the formulas that if $f$ is a one
dimensional solution of these equations,
\begin{equation}
F(x)=f(x^1+\cdots +x^n)
\label{eq:A6}
\end{equation}
is also a solution of the $n$-dimensional problem. Let us show that
this is, in fact, the general solution. Applying $\partial^2/\partial
y^i\partial z^j$ at $y=z=0$ to the Jacobi identity in
eq.~(\ref{eq:A5}) yields
\begin{equation}
0=-F(x)F_{ij}(0)+F_i(0)F_j(-x)-F_j(0)F_i(x)\,,
\end{equation}
that is, $F_i(x)=F_j(x)$ for all $i,j$. This implies
eq.~(\ref{eq:A6}). (This can be seen in the two dimensional case by
making a change of variables to $x^1+x^2$ and $x^1-x^2$, and by
induction for the higher dimensional cases.)

The form $F(x)=f(x^1+\cdots+x^n)$ is already incompatible with the
axioms, since the classical sector requires $f(x)=x$ whereas the
quantum sector requires $f(x)=2\sin(x/2)$. In order to find the
general form of $f(x)$ it is sufficient take the first derivative on
$z$ at $z=x$ and then the first derivative on $y$ at $y=0$ in the
Jacobi identity for $f$. This yields
\begin{equation}
0=f(x)f''(x)+f'^2(0)-f'^2(x)\,,
\end{equation}
which together with $f(-x)=-f(x)$ and $f'(0)=1$ yields
$f(x)=\frac{1}{h}\sin(hx)$ or $f(x)=x$.

\end{document}